\documentstyle[twoside,fleqn,espcrc2]{article}
\input{psfig}
\input{epsf}

\def\mev{\,{\rm Me\kern-0.1em V}}
\def\gev{\,{\rm Ge\kern-0.1em V}}

\title{Matching the High Momentum Modes in a Truncated Determinant Algorithm}

\author{A.~Duncan\address{Dept. of Physics and Astronomy, Univ. of Pittsburgh,
        Pittsburgh, PA 15260},%
        E.~Eichten\thanks{Presenter}\address{Fermilab, P.O. Box 500, Batavia, IL 60510}%
        ~and 
        H.~Thacker\address{Dept.of Physics, University of Virginia,
        Charlottesville, VA 22901}} 

\begin{document}

\begin{abstract}
Within a truncated determinant algorithm,
two alternatives are discussed for including systematically
the remaining ultraviolet modes. Evidence is presented that
these modes are accurately described by an effective action involving
only small Wilson loops.
\end{abstract}

\maketitle

\section{Introduction}

Because QCD in four dimensions is renormalizable, not 
super-renormalizable, the fluctuations of 
the fermion determinant are significant at all physical scales.  
Fortunately the short distance behaviour of QCD is very well understood.
In particular, we know that for sufficiently high momentum scales 
this physics should be accurately described by an improved gauge action.

In the truncated determinant algorithm\cite{tda98},
the fermion determinant is separated into two pieces
\begin{equation}
\ln {\rm det} H = [{\rm Tr} \ln H]_{low \;\lambda} + [{\rm Tr} \ln H]_{high \;\lambda} 
\end{equation}
where the lowest $n_{\rm cut}$ eigenvalues are directly calculated 
and included in the Monte Carlo updating procedure. The higher eigenvalues 
can be included in the
Monte Carlo by some approximation that matches onto the 
low eigenvalue results without gaps or double counting, 
is controlled  and becomes exact in the
continuum limit. 

Two numerical methods suggest themselves for  calculating the 
high eigenvalues: 
(1)The multiboson approach of L\"{u}scher\cite{multiboson}.
(2)Using a small number of gauge loops to model the determinant as proposed
by  Sexton and Weingarten \cite{Wein97}, and Irving and Sexton \cite{Irving97}.

\section{Matching onto the Multiboson Method}

One method to compute the high eigenvalues which is guaranteed to succeed
is the multiboson approach of L\"{u}scher\cite{multiboson}. 
Define 
\begin{equation}
P_{eff}(U) \equiv [{\rm det(D+m)}]^{n_f} \exp{(-S_g(U))} 
\end{equation}
and 
\begin{equation}
H = \gamma_5 {\rm(D+m)}/[{\rm c_m(8+m)}]\;\;\; (c_m \geq 1) 
\end{equation}
where $c_m$ is chosen so that the eigenvalues of $H$ are in the interval $(-1,1)$.
L\"{u}scher chooses a sequence of polynomials $P_{n}(s)$ of even degree  n such that
\begin{equation}
\lim_{n\rightarrow \infty} P_{n}(s) = 1/s \;\;\;{\rm for \;\;all\;\;} 0 < s \leq 1
\end{equation}
then for $n_f = 2$
\begin{equation}
 {\rm det} H^2 = \lim_{n\rightarrow \infty}[ {\rm det} P_{n}(H^2)]^{-1}
\end{equation}
Choose polynomials such that complex roots $z_1 \dots z_n$ come in complex
conjugate pairs (non real) so that  $\sqrt{z} = \mu + i \nu$.
Then 
\begin{equation}
 {\rm det} H^2 = \lim_{n\rightarrow \infty}\prod_{k=1}^n {\rm det}[(H-\mu_k)^2+\nu_k^2]^{-1}
\end{equation}
Hence we can write 
\begin{equation}
 P_{eff}(U) = \lim_{n\rightarrow \infty}\frac{1}{Z_b} \int D\phi D\phi^{\dagger}
 \exp{-(S_g+S_b)}
\end{equation}
where the bosonic action is given by 
\begin{equation}
S_b = \sum_{k=1}^n \sum_{x}|(H-\mu_k)\phi_k(x)|^2 + \nu^2|\phi_k(x)|^2 .
\end{equation}
L\"{u}scher used Chebyshev polynomials
to estimate how many boson fields ($n_b$) are required to represent the original action
to a fixed accuracy in the range ($\epsilon < s \leq 1$) 
The error is given by\cite{multiboson}:
\begin{equation}
|R(s)| \leq 2 (\frac{1-\sqrt{\epsilon}}{1+\sqrt{\epsilon}})^{n_b+1}.
\end{equation}
Therefore, the convergence is exponential with rate $2\sqrt{\epsilon}$
as $n_b\rightarrow \infty$. 

A practical problem with this 
multiboson method is that it requires an increasingly large number of boson 
fields as the quark mass becomes lighter. As $m_q \rightarrow 0$, we must 
take $\epsilon \rightarrow 0$, but to obtain a fixed level of accuracy 
we must hold $2\sqrt\epsilon n_b$ fixed and hence $n_b$ increases without bound.

However the multiboson method matches nicely onto the calculation of 
low eigenvalues. This was first suggested by Alexandrou et.al.\cite{addlowmb}.
In the truncated determinant method, the cutoff $\epsilon$ for the
multiboson method is set by the highest eigenvalue of $H^2$
which is explicitly included in the low end calculations.  Hence 
it does not explode as the quark mass goes to zero. The combination of 
methods remain accurate for all quark masses. For example, for $\beta$ = 5.9
on a 12$^3$x24 lattice with direct inclusion of the lowest 100 eigenvalues,
the associated cutoff for the multiboson simulation of the high eigenvalues
is $\sqrt{\epsilon} \approx$ 0.035 independent of the light quark mass.

Furthermore, the error associated
with the inaccurate behaviour of the polynomial fit in the 
range $0 < s < \epsilon$ can be corrected as low eigenvalues are computed
for every configuration update.  We obtain a reweighting term,
\begin{equation}
 \Delta S_b =  \sum_{i=1}^{n_{\rm cut}}\ln {(\lambda_i^2  P(\lambda_i^2))}
\end{equation}
which can be included to eliminate errors in the region $0 < s \leq \epsilon$.

\section{Matching onto Small Loops}

\begin{figure}
\psfig{figure=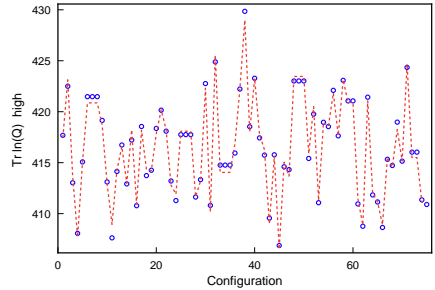,
width=0.80\hsize}
\vspace*{-.2in}
\caption{Comparison of $({\rm Tr} \ln H)_{{\rm high}\lambda}$ (dashed line)
and best fit (up to 6 links) effective gauge action (open circles) for 75 gauge
configurations.}
\label{fig:det_hilo}
\end{figure}

Using the multiboson method for the high end of the determinant satisfies 
all our requirements and completes the algorithm. 
However it may be possible to reduce the total required 
computations even further using a more physical approach to the high eigenvalues.
Consider how many of the high eigenvalues actually
have physical information and are not just lattice artifacts. For example,
for a 12$^3$x24 lattice with $\beta$ = 5.9 and $\kappa =$ .1587 there 
are 497,664 total eigenvalues of the Wilson-Dirac operator; while for a 
high energy cutoff of $1 \gev$ we have approximately 1500 eigenvalues (0.3\%).  
For a fixed volume V and quark mass $m_q$ a decreasing fraction of the eigenvalues
are below a fixed physical scale as $\beta \rightarrow \infty$. Therefore, most of the range of large s fit in L\"{u}scher's  multiboson method
is  physically unimportant. 

This suggests a more physically motivated method for dealing with 
the high eigenvalue part of the fermion determinant 
in which one approximates the ultraviolet contribution to
the quark determinant with an effective gauge action: 
\begin{equation}
[{\rm Tr} \ln H]_{high \;\;\lambda} \approx \sum_{i=0}^{i_{max}} \alpha_i L_i
\end{equation}
where each $L_i$ is a set of gauge links which form a closed path. The 
natural expansion is in the number of links. 
For zero links $L_0$ is just a constant, for four links we have a plaquette, and six
links give the three terms found in considerations of improved gauge 
actions \cite{improvact}.

This idea was studied in detail by Irving and Sexton\cite{Irving97}. These 
studies were done on a 6$^4$ lattice at $\beta$ = 5.7 with 
Hybrid Monte-Carlo full QCD 
simulations (with a heavy sea quark). Their results were rather discouraging. 
It was hard to get a good approximation to the determinant 
with a closed set of loops 
and they needed large loops to even approach a reasonable 
fit\cite{Irving97}.

There are however two important differences between their study 
and our situation.
First, they simulated the {\em whole determinant}, while here we only need to
approximate the eigenvalues above some cutoff. Hence  
we would expect the small loops to dominate at least for sufficiently high
cutoff.
Second, they used an approximate procedure to  estimate stochastically
the logarithm of the determinant needed,
while we are exactly computing all eigenvalues for this study.
It turns out that these differences are critical, as using 
approximately the same lattices 
(and with even lighter quarks) we find an excellent 
approximation to the high end with only small loops. 

We generated a set of 75 configurations on a 6$^4$ lattice  
at $\beta$ = 5.7 and $\kappa =$ .1685. We included the lowest 30 eigenvalues
(which corresponds to a physical cutoff of approximately $\simeq$350 Mev)
in the Monte Carlo accept/reject step in the generation of these 
independent configurations.

Considering only the high eigenvalues, an excellent fit to the fluctuations
is obtained including four and six link closed loops. The variance of the
fit is 0.265. The comparison between the fluctuations in the exact ($S_t$) and
approximate ($S_a$) actions for the high eigenvalue piece is shown in Fig \ref{fig:det_hilo}.
As expected, if only the plaquette term is included the 
variance is larger (2.25) and we must move the low eigenvalue cutoff
to $N=50$ ($\approx$ 700 MeV) to reduce the variance below one.
The results for various cutoffs and terms included are shown in 
Table 1.

\begin{table}
\begin{center}
\begin{tabular}{|c|c|c|c|}
\hline
\multicolumn{1}{|c|}{$n_{\lambda}$}
&\multicolumn{1}{c|}{$\lambda$}
&\multicolumn{1}{c|}{4 links}
&\multicolumn{1}{c|}{6 links} \\
 cut & (MeV) &  & \multicolumn{1}{c|}{(with WL)} \\ 
\hline
0 & 0 & 4.98 & 1.074 (0.835) \\
$\pm 15$ & 340 & 2.25 & 0.2652 (0.233) \\
$\pm 50$ & 700 & 0.940 & 0.0564 (0.0491) \\
$\pm 250$ & 1,210 & 0.0733 & 0.0695 (0.0641) \\
$\pm 1250$ & 2,220 & 0.138 & 0.0198 (0.0180) \\
\hline
\end{tabular}
\label{tbl:chisq}
\caption{Variance ($<(S_a - S_t)^2>^{1/2}$) of fit to  high eigenvalues 
of the quark determinant by various sets of
small gauge loops (WL denotes a Wilson line).} 
\end{center}
\end{table}

The linear combination (.46,-.55,.04,.70,.03) 
for (plaquette,rectangle,chair,polygon,wilsonline) gives the best fit to the high 
eigenvalues of the quark determinant (with $n_{\rm cut} = 30$). 
The configuration to configuration variations of the individual 6-link terms are 
highly correlated. This is to be expected since these three terms are not independent. 

The coefficients of the effective action should be 
independent of the physical volume
with other physical parameters held constant. We are repeating this study for 
an ($8^4$) lattice. The preliminary study on 41 configurations gives the same fit 
parameters within the statistical accuracy. With $n_{\rm cut} = 120$
($\lambda \approx 350 \mev$) the variance is 0.52. 

Although more study is required, this second method looks very attractive
for dealing with the high end of the fermion determinant in full 
QCD with light dynamical quarks. 
Simulations would be performed by
including the predetermined effective gauge action $S_a$ in the
gauge updates and computing the infrared part of the determinant as
in the truncated determinant simulations. 




\begin{thebibliography}{99}

\bibitem{tda98} A. Duncan, E. Eichten and H. Thacker, hep-lat 9806020; 
A. Duncan, these proceedings.

\bibitem{multiboson} M. L\"{u}scher, Nucl. Phys. {\bf B418}, 637 (1994).

\bibitem{Wein97} J.~C.~Sexton and D.~H.~Weingarten, 
Phys. Rev. {\bf D55}  4025 (1997).

\bibitem{Irving97} A.~C.~Irving and J.~C.~Sexton, 
Phys. Rev. {\bf D55}  5456 (1997);
A.C. Irving, J.C. Sexton and E. Cahill,hep-lat/9708004.

\bibitem{addlowmb} C.~Alexandrou, A.~Borrelli, Ph. de Forcrand, A.~Galli,
and F.~Jegerlehner, Nucl. Phys. {\bf B456}, 296 (1995).

\bibitem{improvact} M. L\"{u}scher and P. Weisz, 
Phys. Lett. {\bf 158B} 250 (1985), and references therein.

\end{thebibliography}
\end{document}